\documentclass{aa}  
\usepackage[varg]{txfonts}
\usepackage{graphicx}
\usepackage{amssymb,amsmath}
\usepackage{booktabs,multirow}
\usepackage{bm}
\usepackage{natbib}
\bibpunct{(}{)}{;}{a}{}{,}

\begin{document}

\title{Magnetic helicity and eruptivity in active region 12673}

\author{K. Moraitis$^1$ \and X. Sun$^2$ \and \'E. Pariat$^1$ \and L. Linan$^1$}

\institute{LESIA, Observatoire de Paris, Universit\'{e} PSL, CNRS, Sorbonne Universit\'{e}, Universit\'{e} de Paris, 5 place Jules Janssen, 92195 Meudon, France \and Institute for Astronomy, University of Hawaii at Manoa, Pukalani, HI 96768-8288, USA}

\date{Received ... / Accepted ...}

\abstract{In September 2017 the largest X-class flare of Solar Cycle 24 occured from the most active region (AR) of this cycle, AR 12673. The AR attracted much interest because of its unique morphological and evolution characteristics. Among the parameters examined in the AR was magnetic helicity, but either only approximately, and/or intermittently.}{This work is interested in studying the evolution of the relative magnetic helicity and of the two components of its decomposition, the non-potential, and the volume-threading one, in the time interval around the highest activity of AR 12673. Special emphasis is given on the study of the ratio of the non-potential to total helicity, that was recently proposed as an indicator of ARs eruptivity.}{For these, we first approximate the coronal magnetic field of the AR with two different optimization-based extrapolation procedures, and choose the one that produces the most reliable helicity value at each instant. Moreover, in one of these methods, we weight the optimization by the uncertainty estimates derived from the Helioseismic and Magnetic Imager (HMI) instrument, for the first time. We then follow an accurate method to compute all quantities of interest.}{The first observational determination of the evolution of the non-potential to total helicity ratio seems to confirm the quality it has in indicating eruptivity. This ratio increases before the major flares of AR 12673, and afterwards it relaxes to smaller values. Additionally, the evolution patterns of the various helicity, and energy budgets of AR 12673 are discussed and compared with other works.}{}

\keywords{Sun: flares -- Sun: activity -- Magnetohydrodynamics (MHD) -- Methods: numerical}

\titlerunning{Magnetic helicity and eruptivity in AR 12673}
\authorrunning{Moraitis et al.}

\maketitle

\section{Introduction}
\label{sect:introduction}

Magnetic helicity is a physical quantity that is used often in investigations of the eruptive behaviour of solar active regions \citep[ARs,][]{rust94,nindos04,green18}. This is because it is a conserved quantity of ideal magneto-hydrodynamics (MHD), and thus it determines the dynamics of magnetized systems \citep{taylor74,pariat15,linan18}. Magnetic helicity is a geometrical quantity that describes the level of complexity of a magnetic field, through the twist and writhe of its field lines, and their interlinking.

Magnetic helicity however is well defined only for systems that are magnetically closed, and thus is of limited use in the Sun and all other astrophysical conditions. The appropriate form of helicity in these situations is relative magnetic helicity \citep{BergerF84,fa85}. It is defined through a volume integral as
\begin{equation}
H=\int_V (\mathbf{A}+\mathbf{A}_\mathrm{p})\cdot (\mathbf{B}-\mathbf{B}_\mathrm{p})\,\mathrm{d}V,
\label{eqhr}
\end{equation}
where $\mathbf{B}$ denotes the three-dimensional (3D) magnetic field in the volume of interest, $V$, while $\mathbf{B}_\mathrm{p}$ is a reference field that usually is taken to be potential. The two fields are generated from the vector potentials $\mathbf{A}$ and $\mathbf{A}_\mathrm{p}$. The potential magnetic field is chosen to have the same normal components with $\mathbf{B}$ along the boundary of the volume, $\partial V$, namely
\begin{equation}
\hat{n}\cdot\mathbf{B} \rvert_{\partial V}=\hat{n}\cdot\mathbf{B}_\mathrm{p} \rvert_{\partial V},
\label{eqnormal}
\end{equation}
with $\hat{n}$ denoting the unit vector that is normal to $\partial V$. This choice guaranties that the relative magnetic helicity given by Eq.~\ref{eqhr} is independent from the gauges of the vector potentials, and thus physically meaningful.

Relative magnetic helicity can also be uniquely split into two gauge-independent components \citep{berger99}, namely
\begin{equation}
H=H_\mathrm{j}+H_\mathrm{pj}.
\end{equation}
These are the non-potential component
\begin{equation}
H_\mathrm{j}=\int_V (\mathbf{A}-\mathbf{A}_\mathrm{p})\cdot (\mathbf{B}-\mathbf{B}_\mathrm{p})\,\mathrm{d}V,
\label{eqhj}
\end{equation}
that depends only on the current-carrying part of the magnetic field, $\mathbf{B}_\mathrm{j}=\mathbf{B}-\mathbf{B}_\mathrm{p}$, and the volume-threading component
\begin{equation}
H_\mathrm{pj}=2\int_V \mathbf{A}_\mathrm{p}\cdot (\mathbf{B}-\mathbf{B}_\mathrm{p})\,\mathrm{d}V,
\label{eqhpj}
\end{equation}
that depends additionally on the potential field. The detailed analysis of the behaviour of these components, and of their dynamics in \citet{linan18}, showed that, unlike relative magnetic helicity, they are not conserved quantities of ideal MHD.

The study of the different components of relative magnetic helicity is important however, as they provide additional information compared to $H$. Already from the first separate examination of these two components in eruptive ARs \citep{moraitis14}, it was observed that the non-potential component was fluctuating in accordance with the eruptions of the AR. This was noted in both observed, and synthetic, MHD-modelled eruptive ARs.

Moreover, using this decomposition of helicity in a set of eruptive, and non-eruptive MHD flux-emergence simulations, \citet{pariat17} have determined that the ratio $\lvert H_\mathrm{j}\rvert/\lvert H\rvert$ is an excellent indicator of the system's eruptivity. They noticed that the helicity ratio behaved differently in the eruptive than in the non-eruptive simulations, as well as between the pre-, and post-eruptive phases of the former. The helicity ratio obtained high values only in the eruptive cases, and only before the eruptions. The possible importance of $\lvert H_\mathrm{j}\rvert/\lvert H\rvert$ was also highlighted in \citet{linan18} by using an MHD simulation of the formation of a coronal jet. Again, the helicity ratio attained very high values during the generation of the jet, and it droped significantly after that.

This result was further tested in \citet{zuccarello18} with a different set of line-tied, eruptive MHD simulations where the quality of the helicity ratio as an eruptivity indicator was reconfirmed. Additionally, by carefully identifying the onset of the eruption in these simulations, it was found that a threshold value of $\lvert H_\mathrm{j}\rvert/\lvert H\rvert\simeq 0.3$ was reached by all simulations when the eruptions occured.

These promising results on the ratio $\lvert H_\mathrm{j}\rvert/\lvert H\rvert$ led to an interest in deriving it in observations. In a first observational determination of the helicity ratio, \citet{james18} have found the value $\lvert H_\mathrm{j}\rvert/\lvert H\rvert\simeq 0.17$ for the flux rope of an AR, one hour before the occurence of an eruption. A reasonable next step is to study the evolution of the helicity ratio in an observed AR, which is the central task of this work.

\begin{figure}[ht]
\centering
\includegraphics[width=0.44\textwidth]{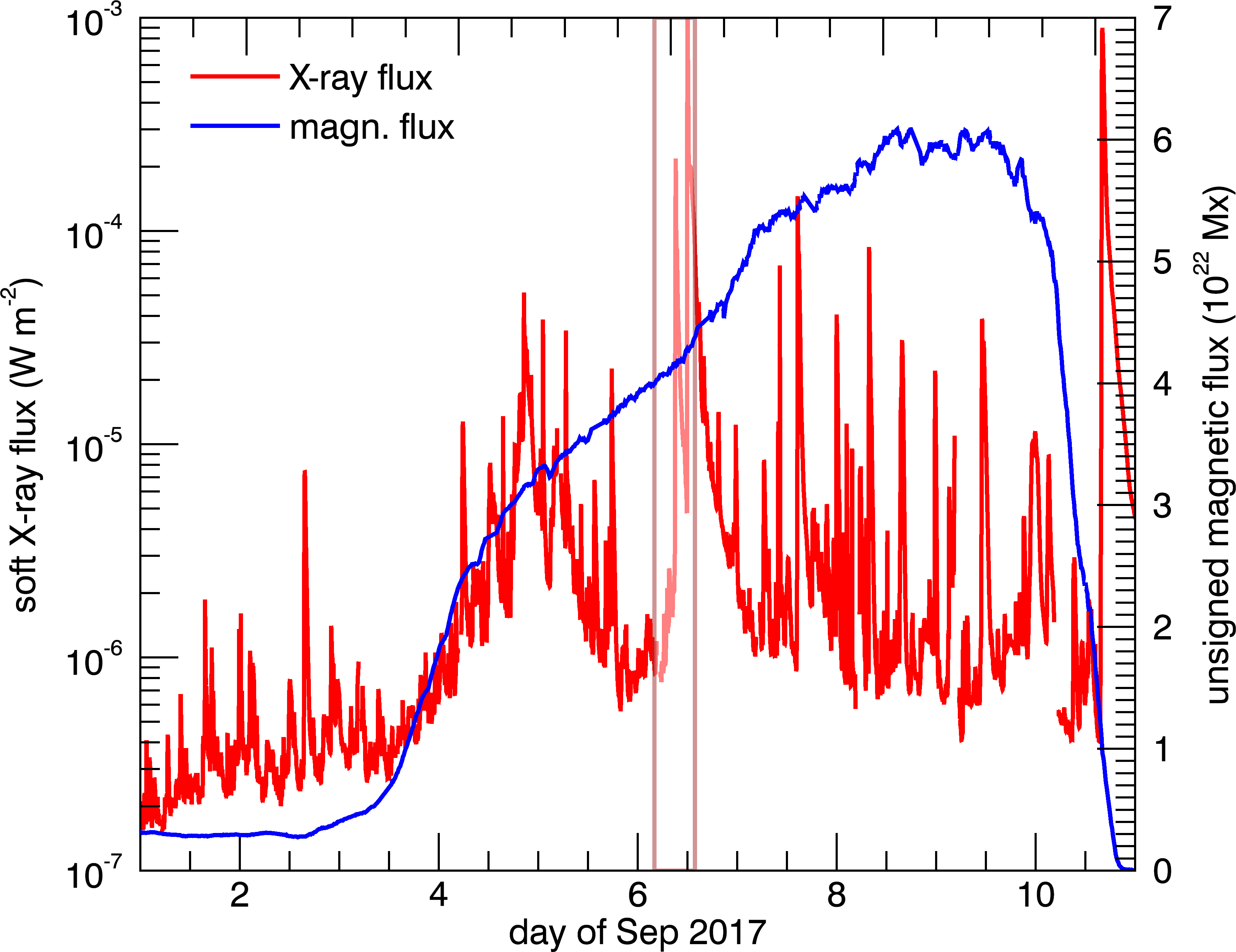}\\
\includegraphics[width=0.44\textwidth]{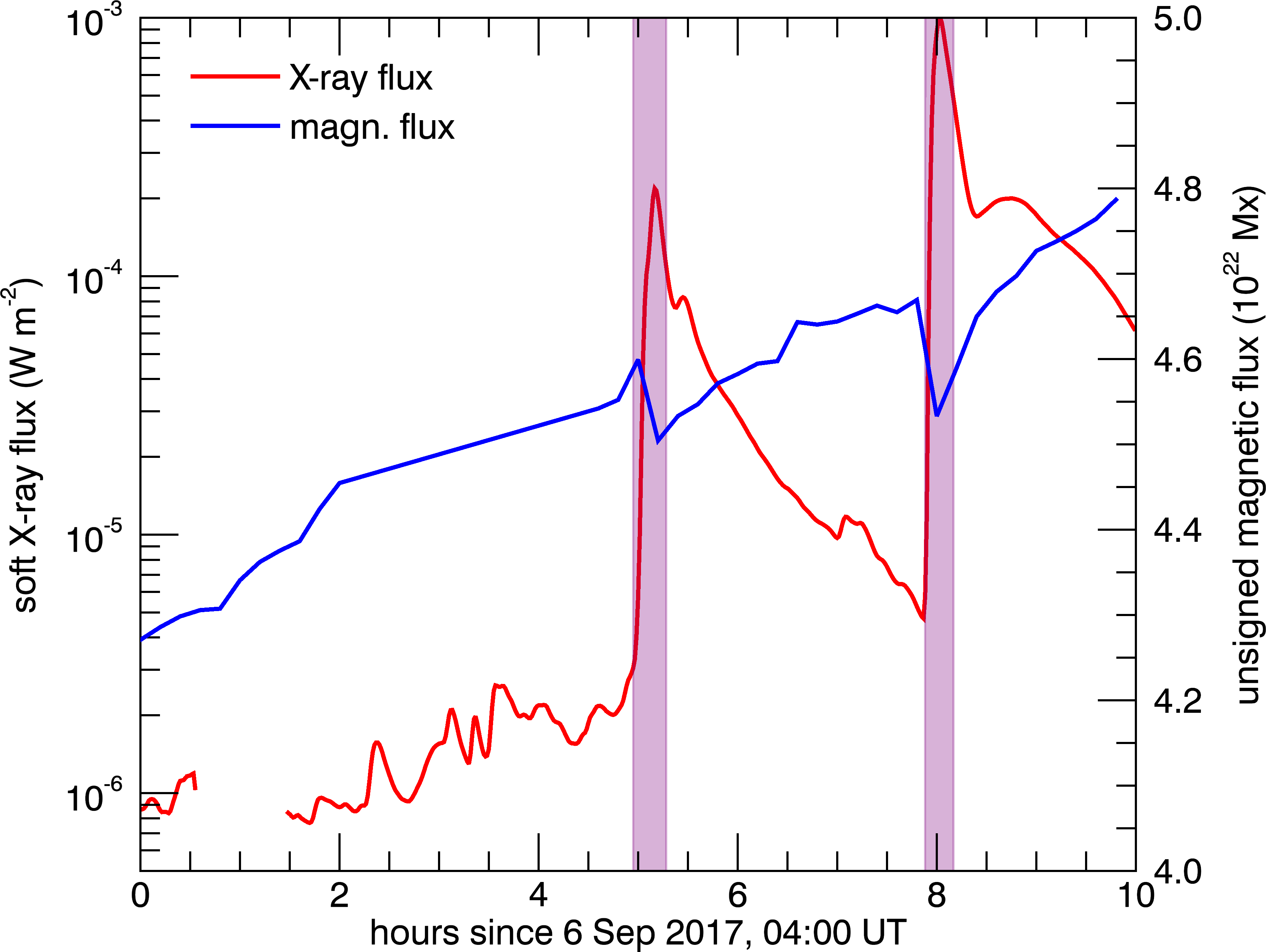}
\caption{GOES soft X-ray light curve, and evolution of HMI unsigned magnetic flux in AR 12673, for the first ten days of September 2017 (top). The brown vertical lines denote the interval of interest for this study, which is shown zoomed in at the bottom. The purple bands in that plot denote the time intervals of the X2.2 and the X9.3 flares, with onset times at 08:57 UT and 11:53 UT, respectively. The HMI magnetic flux on the bottom plot is of limited accuracy due to the emission of the flares, while between 06:00 UT and 08:36 UT there were no observations due to an eclipse of SDO.}
\label{figgoes}
\end{figure}

A perfect target for this is the most active region of Solar Cycle 24, AR 12673 \citep{sun17}. During its passage through the solar disk on the week of 4-10 September 2017 it produced four X-class flares, including the two strongest of the cycle, 27 M-class flares, and numerous more of smaller size, as the Geostationary Operational Environmental Satellite (GOES) soft X-ray ($0.1-0.8\,\mathrm{nm}$) light curve of Fig.~\ref{figgoes} depicts.

A few of these flares also produced coronal mass ejections (CMEs). Of the three halo CMEs, two were Earth-directed and geoeffective \citep{chertok18}, while one of them led to particle acceleration at the produced shock front \citep{morosan19}, and also to a ground-level enhancement of energetic particles \citep{augusto19}. This particle event originated from an X8.2 flare and its associated CME of 10 September 2017 \citep{veronig18}, when AR 12673 was crossing the west solar limb.

Most of the interest in the activity of AR 12673 was focused however on the two X-class flares of 6 September 2017; the confined X2.2 flare that started at 08:57 UT (SOL2017-09-06T08:57), and the eruptive X9.3 flare, the largest since 2005, that followed three hours after the first, at 11:53 UT (SOL2017-09-06T11:53). In this work, we are interested in a 10-hours time interval centered around the X2.2 flare of 6 September that also includes the X9.3 flare.

The photospheric morphology of AR 12673 started as a single positive-polarity sunspot and quickly became quite complicated, displaying a complex network of $\delta$-sunspots. It exhibited significant flux emergence and the highest instantaneous flux rate ever observed \citep{sun17}, with values reaching up to $10^{21}\,\mathrm{Mx}\,\mathrm{hr}^{-1}$ during the third of September, as Fig.~\ref{figgoes} also shows. Many dipoles emerged close to the initial sunspot and the successive interactions between them resulted in the increased complexity of the system.

The magnetic field of the AR as measured by the Helioseismic and Magnetic Imager \citep[HMI,][]{sche12} instrument onboard the Solar Dynamics Observatory \citep{pes12} was very strong, especially along the polarity inversion line \citep{wang18}. Photospheric shearing and twisting motions were also observed \citep{verma18}, which helped in the creation of the highly non-potential configuration that powered the two flares.

Many models were invoked in order to explain the evolution of AR 12673 towards the two X-class flares and the eruption of the second. \citet{yang17} associated the increased flare-productivity of the AR with the blocking of newly emerging flux by already existing flux, and the eruptive flare with a filament becoming kink unstable. The formation of a coronal sigmoid and the initiation of both X-class flares from the core of the sigmoid, led \citet{mitra18} to characterize the AR as a single ``sigmoid–to–arcade'' event.

\citet{hou18} attributted the generation of the two flares to the formation of multiple flux ropes and twisted loop bundles that eventually erupted. The presence of a large, highly-twisted magnetic flux rope, and some smaller ones, was also confirmed by the MHD modelling of the X9.3 flare of AR 12673 made by \citet{inoue18}.

The highly-twisted flux ropes that were found by many authors indicate that magnetic helicity should be an important factor in the eruptivity of AR 12673. Indeed, \citet{liu18} have found a significant enhancement, by almost a factor of three, of the magnetic helicity during the confined X2.2 flare, although they approximated helicity only with its twist. In another study, \citet{yan18} did not find a significant change of helicity during the first X-class flare, but only a small decrease during the second. They found however a negative helicity injection rate throughout the day of the two flares, consistent with the rotational motions of the sunspots in AR 12673.

The sign of magnetic helicity was actually negative for the whole week of the AR's activity, as was deduced by integrating the helicity injection rate \citep{vemareddy19}. The same author found that AR 12673 exhibited very fast helicity injection, by a factor of three higher compared to other ARs when it was normalized to the square of magnetic flux.

All these support the choice of AR 12673 as an interesting case to study the evolution of the helicity ratio in an observed AR. The question we want to address is whether this evolution relates with the eruptivity of the AR, as proposed by \citet{pariat17}. Apart from $\lvert H_\mathrm{j}\rvert/\lvert H\rvert$, we are also interested in examining the evolution of all helicity-, and energy-related budgets of the AR. For determining all quantites of interest, the coronal magnetic field must first be estimated. The method we followed for this is described in Section~\ref{sect:data}. The results that we obtained are presented in Section~\ref{sect:results}, and they are then discussed in Section~\ref{sect:discussion}.

\section{Method}
\label{sect:data}

This section describes the method we followed to compute all magnetic helicities and energies in AR 12673. We first describe the non-linear force-free (NLFF) field extrapolation method that we used in order to approximate the coronal magnetic field of the AR. Then, we check that the produced coronal fields are suitable for use with the helicity computation method, which we briefly describe afterwards.

\subsection{Coronal field modelling}
\label{sect:nlfff}

\begin{figure}[h]
\centering
\includegraphics[width=0.44\textwidth]{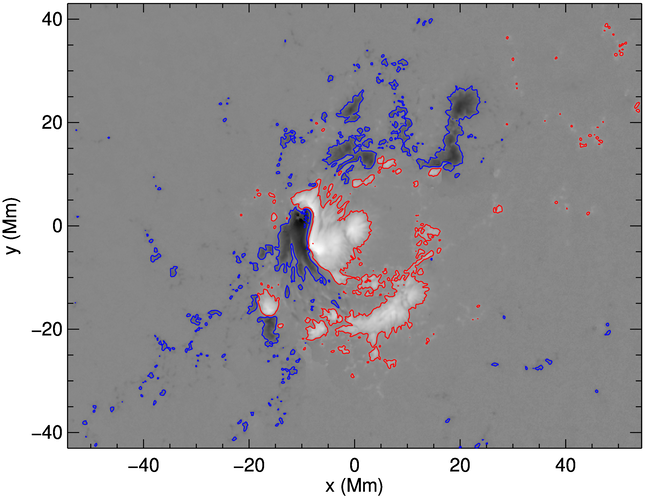}\\
\includegraphics[width=0.44\textwidth]{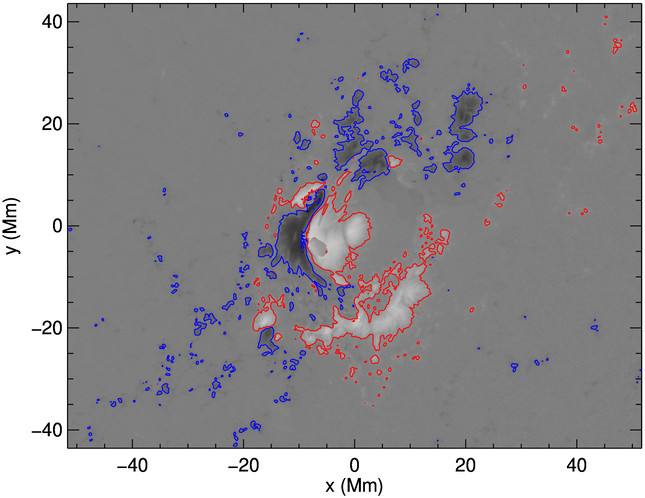}
\caption{Photospheric distribution of $B_z$ in AR 12673 at the beginning, top plot at 04:12 UT, and the end of our study interval, bottom plot at 13:36 UT. Blue contours correspond to $B_z=-500$~G, and red to $B_z=500$~G.}
\label{figbphot}
\end{figure}

The starting point of all computations performed in this work were magnetic field data from the HMI instrument of SDO. We used the vector magnetogram data with 12 minutes cadence (product \texttt{hmi.sharp\_720s}). The actual cadence of the data varied during the interval of interest due to an eclipse of the observatory and a few low-quality data during the two flares. The morphology of the $B_z$-distribution on the photosphere at the beginning and at the end of the 10-hours interval of study is shown in Fig.~\ref{figbphot}.

The next step was to model the 3D magnetic field of AR 12673 in a finite volume above the AR with a NLFF field extrapolation method. Such methods try to simultaneously satisfy the force-free, and the solenoidal conditions for the magnetic field, which read
\begin{eqnarray}
\label{eqnlff}
(\nabla\times\mathbf{B})\times\mathbf{B}&=&0\label{eqnlffa}
\\
\nabla\cdot\mathbf{B}&=&0.
\label{eqnlffb}
\end{eqnarray}
The unavoidable presence of numerical errors in the solution process hinders the exact fullfilment of these conditions. The solenoidal condition is thus never fully constrained in a NLFF extrapolated magnetic field. The level of non-solenoidality is however very important for the helicity computations. \citet{valori16} have showed that the values of helicity are unreliable when the non-solenoidality errors are above a certain level. It is therefore essential to estimate the level of solenoidality of the given magnetic field when computing relative magnetic helicity, which we do with two ways, as described in Sect.~\ref{sect:solenoid}.

A possible way to optimize the 3D magnetic field model with respect to its divergence-freeness is to perform different extrapolations and then to keep the best-performing one. In this work we estimated the magnetic field of AR 12673 using two different NLFF methods. The investigation of the various parameters in each method was not exhaustive since our goal was to have a reliable magnetic field for the helicity computations and not to determine it with the highest possible accuracy.

The first extrapolation method that we used was the optimization method of \citet[][hereafter W04]{wieg04}. This tries to minimize a functional so that to fullfil the divergence-, and force-free conditions, given by Eqs.~\ref{eqnlffa}, \ref{eqnlffb}. We note at this point that the assumption that the magnetic field is force-free is not always true, especially near the photosphere, and during solar flares. For this, the magnetograms were first preprocessed so that to become more compatible with the force-free assumption \citep{wieg06} with the standard set of preprocessing parameters $[\mu_1,\mu_2,\mu_3,\mu_4]=[1,1,10^{-3},10^{-2}]$, in the authors' original notation.

The resolution used in the extrapolations was 1", or 720~km, while the grid size was 320$\times$320$\times$320 pixels. The computational volume consisted of the volume of interest surrounded by a boundary buffer layer in the lateral and top boundaries with 16 pixels on each side. In the helicity computations only the inner, physical part of the fields were kept, after further cutting in height, at roughly the $2/3$ of the total height. The final datacubes thus consisted of 288$\times$288$\times$203 pixels. The total snapshots were 28, covering 10 hours starting from 6 September, 04:00 UT.

The second extrapolation method that we used was the newer version of the same optimization code \citep[][hereafter W12]{wieg12}, which usually performs better. The idea of W12 is to add another term to the functional of W04 so that to take into account the uncertainties in the measurements of the photospheric magnetic field components. This is done by weighting each pixel separately according to its uncertainty. In this study we used the actual HMI measurement uncertainties for the first time. The empirical weighting function that we employed was: 
\begin{equation}
w = \left\lbrace 
\begin{array}{ll}
0.01+0.99\exp \left( - \frac{\sigma_B}{0.03B} \right), & \hbox{pixels with}\,B \geq 200\,\mathrm{G} \\
0.01, & \hbox{pixels with}\,B < 200\,\mathrm{G}
     \end{array}
         \right. .
\end{equation} 
Here $B$ denotes the magnetic field strength at a pixel on the photosphere, and $\sigma_B$ its uncertainty. We assumed a value of 200~G for the typical noise threshold, and 0.03 for the typical value of $\sigma_B/B$. This choice sharply de-emphasizes the contribution of bad measurements in several strong field regions.

\begin{figure}[h]
\centering
\includegraphics[width=0.46\textwidth]{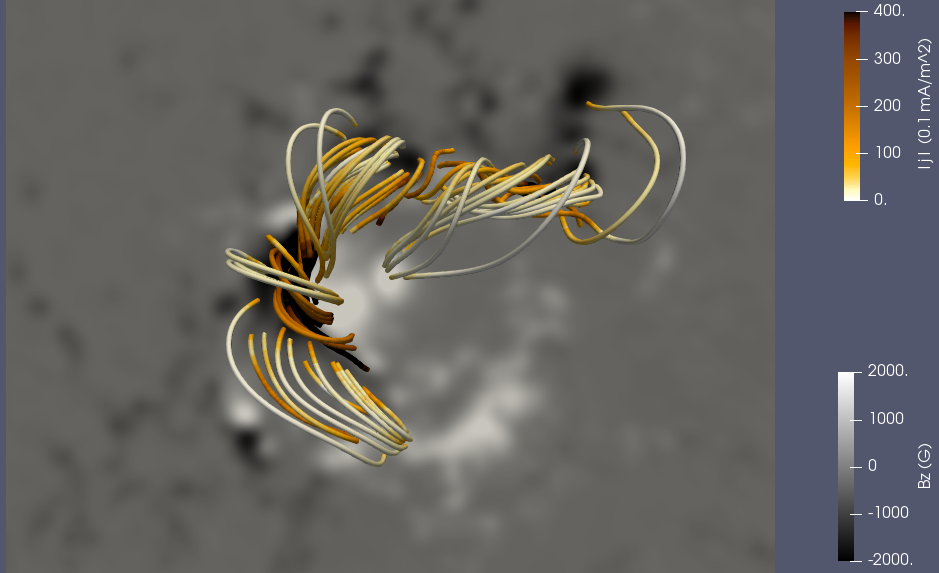}\\[2pt]
\includegraphics[width=0.46\textwidth]{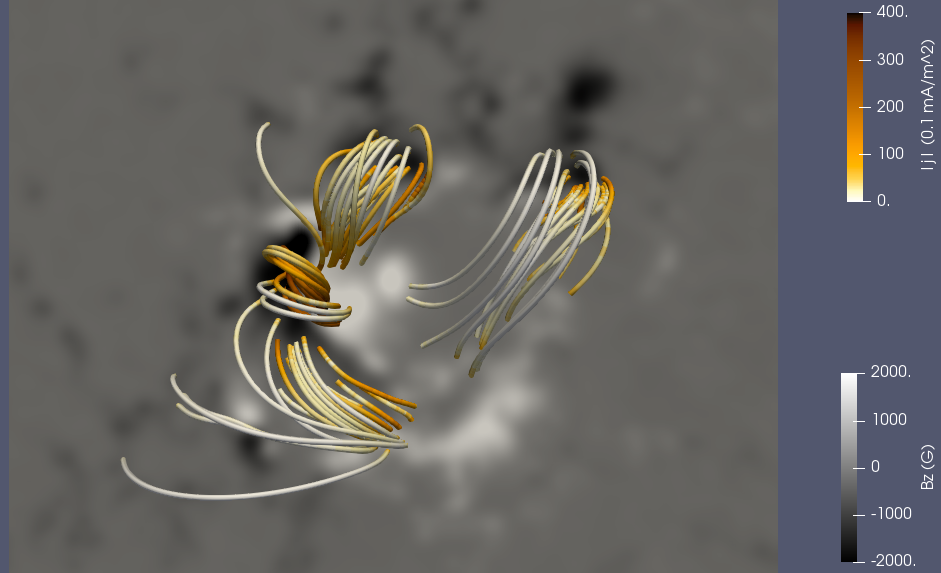}
\caption{Morphology of the reconstructed magnetic field of AR 12673 at 08:48 UT of 6 September 2017 with the two versions of the NLFF extrapolation methods (W04 to the top, W12 to the bottom). The same footpoints were used in both cases. Field lines are colored according to the magnitude of the electrical current density.}
\label{figbfld}
\end{figure}

All remaining parameters were the same as in the W04 case, except from the size of the buffer which was 32 pixels in this case. The final datacubes were thus 256$\times$256$\times$203 pixels. The number of snapshots were also smaller in this case, 18 instead of 28, but with the same coverage until the first flare. The morphology of the 3D magnetic field for the two cases and for the same snapshot is shown in Fig.~\ref{figbfld}. We note that the two cases exhibit many differences in morphology. The morphology of the magnetic field evolves during the flares, following the evolution of the photospheric magnetic field.

We finally note that we checked other cases for the extrapolation parameters also, such as different $\mu_3$ and/or $\mu_4$ pre-processing parameters in W04, different binning in W04, and different weighting of the various terms in the functional of W12. None of these cases produced better results however, and so we presented here the two most accurate cases.

\subsection{Solenoidality level estimation}
\label{sect:solenoid}

The resulting magnetic fields were then tested for their quality with respect to how divergence-, and force-free they were. A traditional way to quantify the divergence-freeness of a magnetic field is through the average absolute fractional flux increase, $\langle|f_i|\rangle$, \citep{wheatland00}. This parameter expresses the average of the local non-solenoidalities in the volume of interest; the smaller it is the more solenoidal the field is. The level of divergence-freeness in the W04 case was quite high with a mean value over all the 28 snapshots of $\langle|f_i|\rangle=(5.9\pm 0.5)\times 10^{-4}$. The respective values for the W12 set were slightly better, with a mean value over all 18 snapshots of $\langle|f_i|\rangle=(5.3\pm 0.6)\times 10^{-4}$.

We also tested the level of force-freeness of the extrapolated fields with an angle, $\theta_J$, that expresses the average current-weighted angle between the current and the magnetic field \citep{wheatland00}. The mean value we found for the W04 magnetic fields was $\theta_J=(16.8\pm 2.6)^\mathrm{o}$, typical for this method. In the W12 case, the force-freeness level was again slightly better, with an average angle $\theta_J=(15.1\pm 1.5)^\mathrm{o}$. 

Another parameter that indicates the divergence-freeness of a magnetic field  is the energy ratio $E_\mathrm{div}/E$, originally used by \citet{valori16}. It expresses the fraction of the total energy that is related to all the (numerical) non-solenoidalities of the magnetic field. To derive the mathematical expresion for $E_\mathrm{div}$ one needs to decompose the magnetic field into potential and current-carrying components, $\mathbf{B}=\mathbf{B}_\mathrm{p}+\mathbf{B}_\mathrm{j}$, and then split each component into solenoidal and non-solenoidal parts, $\mathbf{B}_\mathrm{p}=\mathbf{B}_\mathrm{p,s}+\mathbf{B}_\mathrm{p,ns}$, and $\mathbf{B}_\mathrm{j}=\mathbf{B}_\mathrm{j,s}+\mathbf{B}_\mathrm{j,ns}$, following \citet{val13}. By defining the energy budget of each component through the relation
\begin{equation}
E_\mathrm{x}=\frac{1}{8\pi}\int_V \mathbf{B}_\mathrm{x}^2\,\mathrm{d}V,
\label{eqengen}
\end{equation}
we end up with the following decomposition for the total energy of the given magnetic field
\begin{equation}
E=E_\mathrm{p,s}+E_\mathrm{p,ns}+E_\mathrm{j,s}+E_\mathrm{j,ns}+E_\mathrm{mix},
\label{eqdecomp}
\end{equation}
with $E_\mathrm{mix}$ the energy corresponding to all the cross terms. A non-negative energy associated to all non-solenoidal components of the magnetic field can then be defined by the quantity
\begin{equation}
E_\mathrm{div}=E_\mathrm{p,ns}+E_\mathrm{j,ns}+|E_\mathrm{mix}|.
\label{eqediv}
\end{equation}
It is an upper limit to the non-solenoidal energy since the absolute value of $E_\mathrm{mix}$, the only signed term in the decomposition of Eq.~\ref{eqdecomp}, is considered. A perfectly solenoidal magnetic field has $E_\mathrm{p,ns}=E_\mathrm{j,ns}=E_\mathrm{mix}=0$, and thus also $E_\mathrm{div}=0$; the larger the values of $E_\mathrm{div}$ are, the more non-solenoidal it gets. \citet{valori16} have showed that when $E_\mathrm{div}/E$ exceeds $\sim 8-9\%$, then the values of helicity become unreliable.

\begin{figure}[h]
\centering
\includegraphics[width=0.46\textwidth]{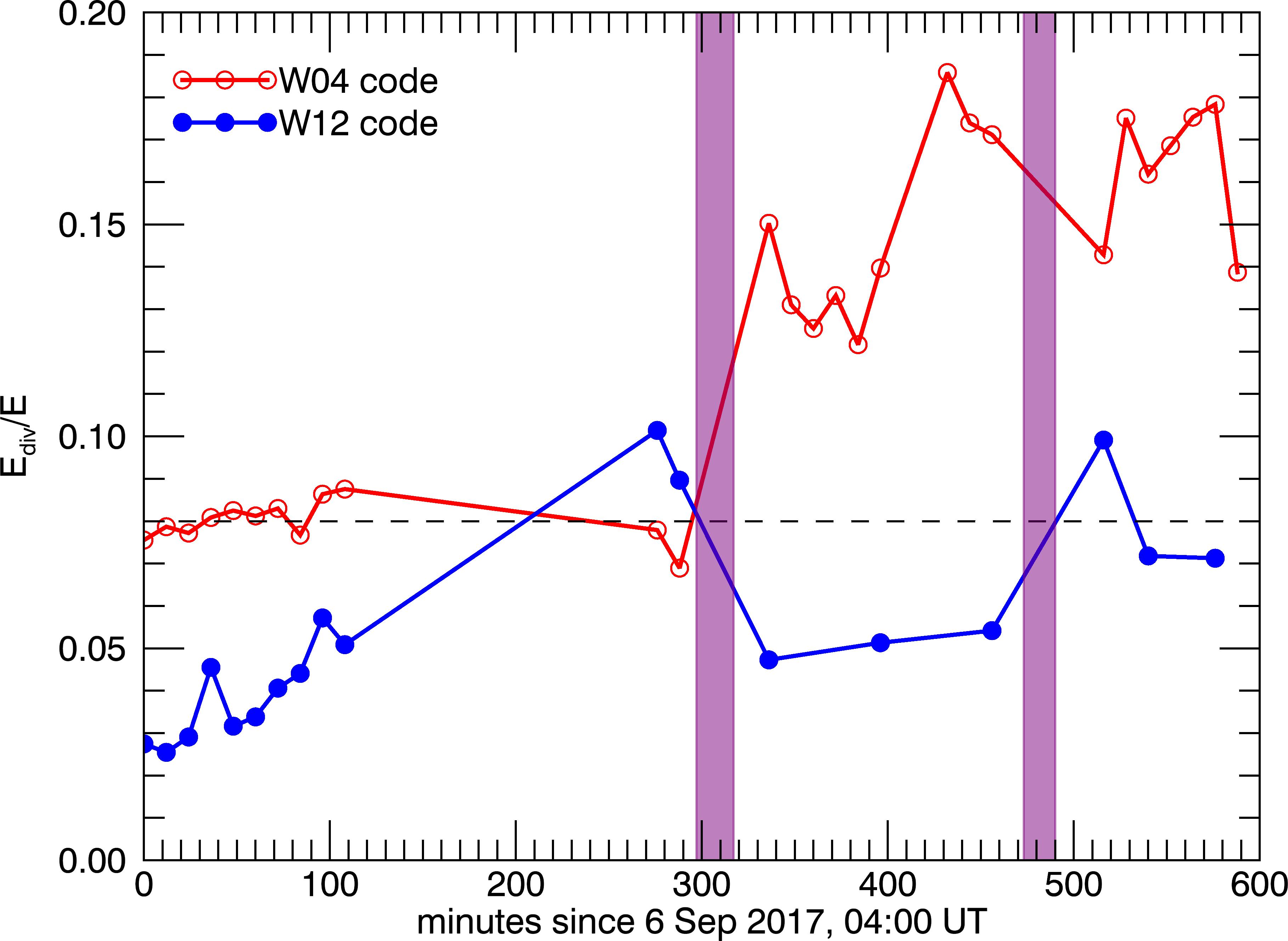}
\caption{Evolution of the divergence-related energy normalized to the total field energy in AR 12673 for 10 hours around the two X-class flares of 6 September 2017, and for the two sets of extrapolated magnetic fields used. The purple bands denote the time intervals of the X2.2 and the X9.3 flares, with onset times at 08:57 UT and 11:53 UT, respectively.}
\label{figediv}
\end{figure}

Since the energy ratio $E_{\rm div}/E$ is important for helicity computations we focused more on this quantity. We calculated all components in the energy decomposition following \citet{val13}, and we show in Fig.~\ref{figediv} the evolution of $E_{\rm div}/E$ for the two sets of extrapolated magnetic fields that we used. Also shown is the condition $E_{\rm div}/E=0.08$ which, according to \cite{valori16}, discriminates between magnetic fields that lead to reliable values of helicity and those that do not.

We note that the energy ratio in the W04 case performs well until the first X-class flare with values around the limit $E_{\rm div}/E=0.08$, but all snapshots after the first flare have much higher values. On the contrary, most of the snapshots in the W12 case are well below the limit, with the exception of three points; the two right before the first flare, and the one after the second flare. Additionally, the values of the energy ratio for the W12 fields are much lower than the W04 ones except from the two points before the first flare. Therefore, we kept in the following computations the best-performing snapshots of the W12 method, except from the two snapshots before the first flare which were replaced by the W04 ones that are better, and the snapshot after the second flare where both methods are above the limit and were thus discarded.

\subsection{Magnetic helicity and energy computations}
\label{sect:enhelc}

For each snapshot in the final dataset we computed relative magnetic helicity from its definition, Eq.~\ref{eqhr}, and also the two gauge-independent components that it splits into, Eqs.~\ref{eqhj}, and \ref{eqhpj}. All the helicities of interest were computed following the method of \citet{moraitis14}. Briefly, this method first calculates the potential magnetic field that satisfies Eq.~\ref{eqnormal} so that the relative magnetic helicity given by Eq.~\ref{eqhr} is gauge independent. The potential field is obtained from the numerical solution of Laplace's equation with pure Neumann boundary conditions. The two vector potentials are then computed following the recipe of \citet{val12}, which makes the clever choice for the gauges, $A_z=A_{\mathrm{p},z}=0$, that results in the straightforward integration of the equations.

We were also interested in the evolution of the various energy budgets of the AR; the total field energy, $E$, the potential energy, $E_\mathrm{p}$, including any non-solenoidal contribution it may have, and the free energy, $E_\mathrm{f}=E-E_\mathrm{p}$. For a perfectly solenoidal magnetic field this free energy coincides with $E_\mathrm{j,s}$ as deduced from Eq.~\ref{eqdecomp}, and deviates from it proportionally to the value of $E_\mathrm{div}$.

\section{Results}
\label{sect:results}

\begin{figure}[h]
\centering
\includegraphics[width=0.49\textwidth]{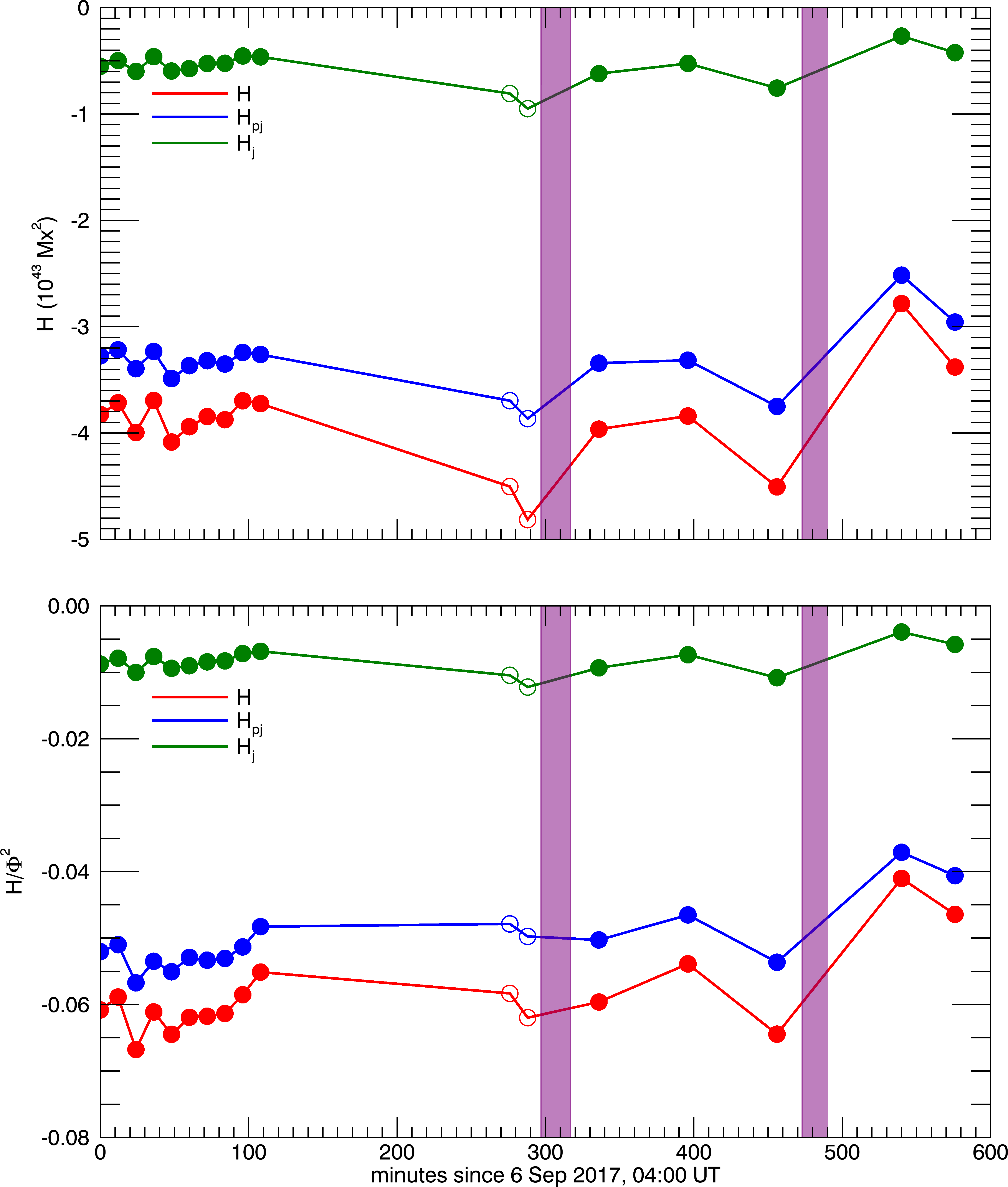}
\caption{Evolution of relative magnetic helicity and its two gauge-independent components in AR 12673 for 10 hours around the two X-class flares of 6 September 2017, in physical units (top panel), and normalized to magnetic flux squared (bottom panel). The purple bands denote the time intervals of the X2.2 and the X9.3 flares, with onset times at 08:57 UT and 11:53 UT, respectively. Filled points correspond to the W12 method and open to the W04 one.}
\label{fighel}
\end{figure}

The evolution of relative magnetic helicity and its two components in AR 12673 is shown in the top panel of Fig.~\ref{fighel}. We notice that relative magnetic helicity is negative, in agreement with previous studies, and that all helicity budgets are also negative. The absolute values of relative magnetic helicity are very high, and reach up to $5\times 10^{43}\,\mathrm{Mx}^2$, mostly because of the high magnetic flux of the AR. The current-carrying component of helicity is much smaller than the other two helicities, as was also found in other cases \citep[e.g.,][]{moraitis14}. The evolution patterns are similar for all helicities, increasing before the first flare, relaxing after it, then again increasing before the second flare and relaxing afterwards. 

The bottom panel of Fig.~\ref{fighel} shows the helicity budgets normalized to the square of the magnetic flux, $\Phi$, which is calculated from the $B_z$ maps of the NLFF fields as $\Phi=\frac{1}{2}\int_\mathrm{phot}|B_z|{\rm d}S$. The normalized helicity exhibits much more typical values compared to the helicity in physical units, confirming that the high values of helicity are due to the high flux. This plot reveals another feature of the helicities evolution. The normalized helicities have a much smoother evolution compared to the regular helicities before the first X-class flare, while afterwards both behave similarly. This could mean that the rise of the helicity budgets before the first flare are related to the increase of the flux then, but the later fluctuations are purely flare-related.

\begin{figure}[h]
\centering
\includegraphics[width=0.46\textwidth]{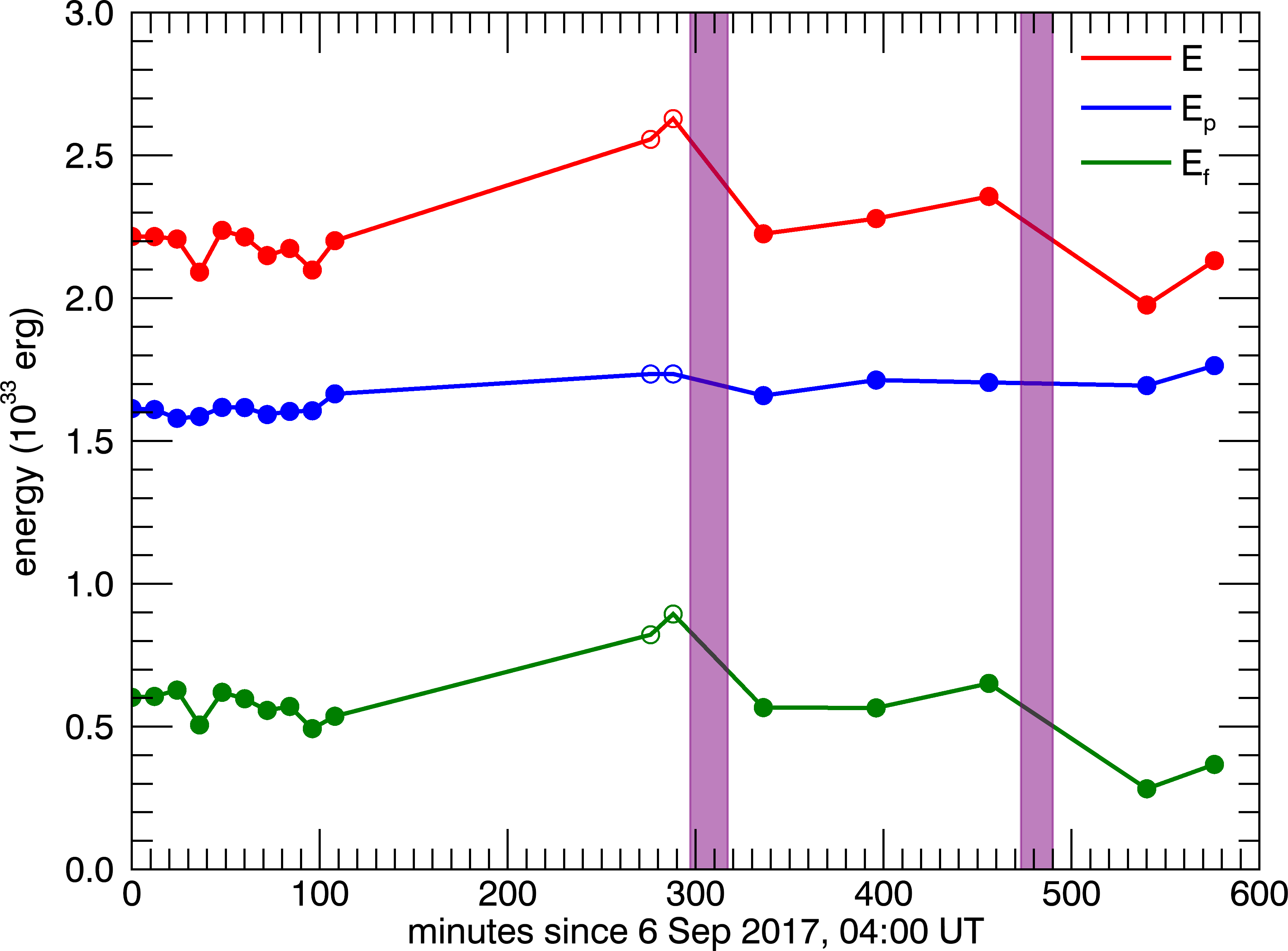}
\caption{Evolution of the various energy budgets in AR 12673 for 10 hours around the two X-class flares of 6 September 2017. The purple bands denote the time intervals of the X2.2 and the X9.3 flares, with onset times at 08:57 UT and 11:53 UT, respectively. Filled points correspond to the W12 method and open to the W04 one.}
\label{fignrg}
\end{figure}

The evolution of the total field energy, the potential energy, and the free energy is shown in Fig.~\ref{fignrg}. Again, all budgets attain very high values due to the high magnetic flux of the AR. We notice that the energy of the potential field changes only slightly during the 10 hours of our study, as a consequence of the modest evolution of the normal field on the boundary that is evident in Fig.~\ref{figbphot}. The change of the normal field is less than that in the horizontal field however, which developed extensive kilogauss bald patches during this period (Sun et al., in preparation). We also note that, although not directly comparable, the evolution pattern of $E_\mathrm{p}$ is quite different that the one of $H_\mathrm{pj}$, despite that both quantities involve the potential field. As a result of the small change of the potential energy, free energy and total field energy show very similar evolution patterns. Moreover, this pattern resembles the one of the helicities, increasing before the flares, and relaxing afterwards.

\begin{figure}[h]
\centering
\includegraphics[width=0.46\textwidth]{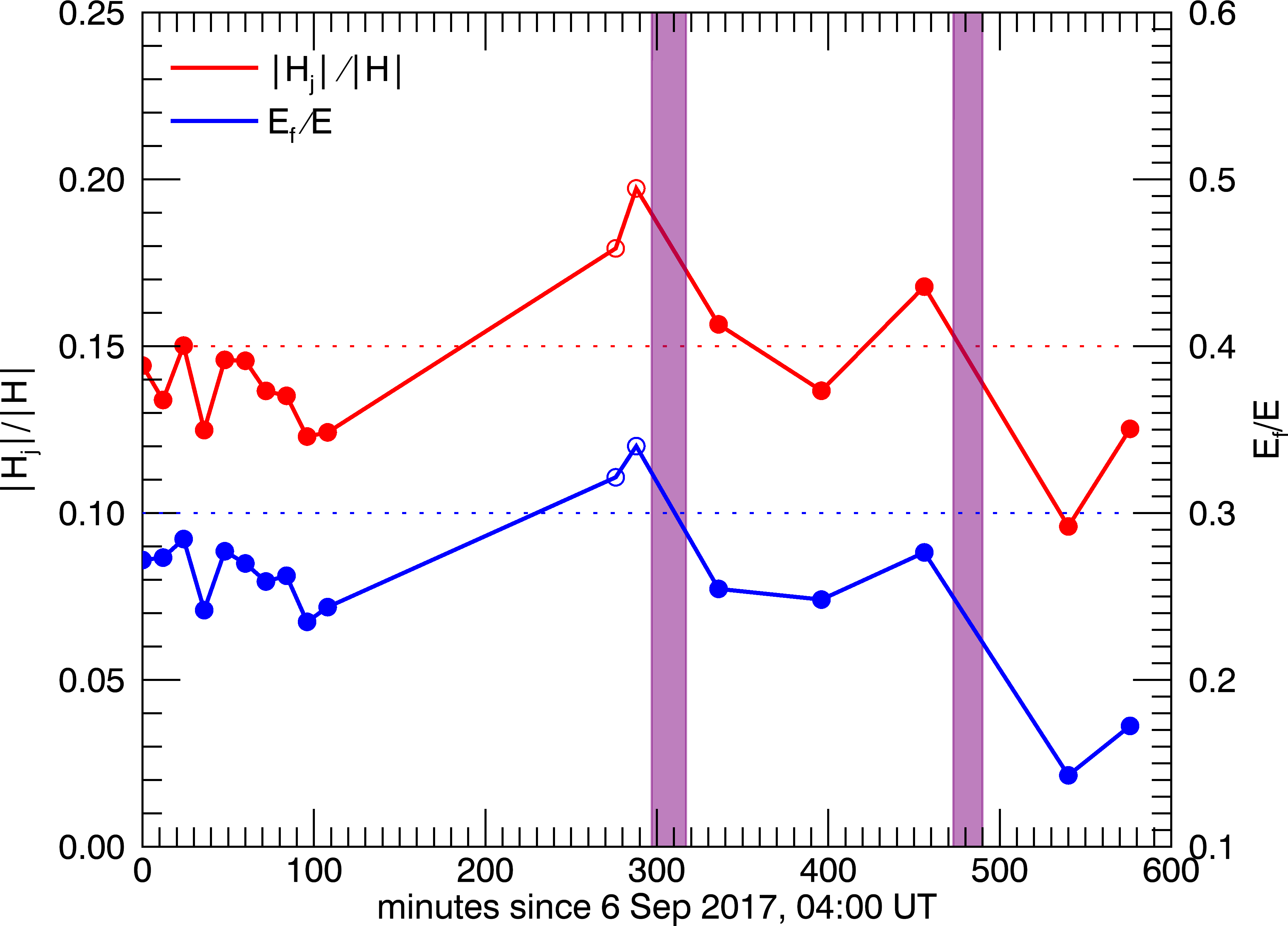}
\caption{Evolution of the ratios $\lvert H_\mathrm{j}\rvert/\lvert H\rvert$ and $E_\mathrm{f}/E$ in AR 12673 for 10 hours around the two X-class flares of 6 September 2017. The purple bands denote the time intervals of the X2.2 and the X9.3 flares, with onset times at 08:57 UT and 11:53 UT, respectively. Filled points correspond to the W12 method and open to the W04 one.}
\label{fighjh}
\end{figure}

The main purpose of this work was to examine whether an observed eruptive AR is compatible with the results of \citet{pariat17} and \citet{zuccarello18}, that the helicity ratio $\lvert H_\mathrm{j}\rvert/\lvert H\rvert$ can indicate AR eruptivity. The evolution of this quantity for AR 12673 is shown in Fig.~\ref{fighjh}. We note that $\lvert H_\mathrm{j}\rvert/\lvert H\rvert$ increases before the two X-class flares and drops after them. The maximum values of the helicity ratio are $\lvert H_\mathrm{j}\rvert/\lvert H\rvert\simeq 0.20$ right before the first X-flare, and $\lvert H_\mathrm{j}\rvert/\lvert H\rvert\simeq 0.17$ before the second. Additionally, these two local maxima are the highest values in the helicity ratio time variation. In other words, a threshold in the value of $\lvert H_\mathrm{j}\rvert/\lvert H\rvert$ can be identified from Fig.~\ref{fighjh} (red dotted line) above which flares occur. Of course, the exact value of this threshold could depend on the conditions of the specific AR and it should be further examined whether a universal threshold exists or not. This result seems to be in agreement with the finding of \citet{pariat17} that $\lvert H_\mathrm{j}\rvert/\lvert H\rvert$ attains its highest values before eruptions. Moreover, the ratio $\lvert H_\mathrm{j}\rvert/\lvert H\rvert$ rises again at the end of our study interval, possibly indicating the subsequent activity of AR 12673. 

Figure~\ref{fighjh} also displays the evolution of the ratio of free energy to total energy, $E_\mathrm{f}/E$, in AR 12673. This parameter has also been examined by \citet{pariat17} and \citet{zuccarello18}, and it was found to be less indicative of eruptivity compared to $\lvert H_\mathrm{j}\rvert/\lvert H\rvert$. The energy ratio evolution pattern that we find exhibits some similarities to the helicity ratio one, which could be a result of the relation between free energy and relative magnetic helicity magnitude \citep{tgr12,tziotziou14}. The local maximum of the energy ratio before the second flare is however much weaker compared to the corresponding local maximum of the helicity ratio. We note in Fig.~\ref{fighjh} that free energy, from the peak of $\sim 34\%$ of the total energy before the first flare, it drops to $\sim 25\%$ between the two flares, and finally to $\sim 15\%$ after the second. Differently than the helicity ratio therefore, the energy ratio cannot be used to set a threshold above which flares occur.

\section{Discussion}
\label{sect:discussion}

This paper dealt with the detailed study of the helicity and energy properties of AR 12673 at the time around its two consecutive X-class flares of 6 September 2017. All involved quantities were accurately computed for the whole period of intense activity, for the first time in such detail. The computations followed a recent finite volume method, and the careful modelling of the 3D magnetic field of the AR.

In the latter, the coronal magnetic field was extrapolated with two different versions of an optimization-based NLFF method. In the newer version of the method, an observationally-derived, empirical weighting that employs the HMI uncertainty estimate was used, for the first time on such methods. This version turned out to be the most accurate, as deduced by a parameter that discriminates extrapolations leading to reliable helicity values from those that do not, but not in all the cases examined. The older version performed better in two snapshots right before the first X-class flare. The best-performing NLFF fields from the two methods were compiled into the final dataset used.

In a very recent work \citep{thalmann19} the authors suggest that the threshold in the parameter $E_\mathrm{div}/E$ should be lower that the one used in this work, $5\%$ instead of $8\%$, for a NLFF field to be used in helicity studies. We expect to see differences of the order of $4-5\%$ to the derived helicities with such a NLFF field, as follows from \citet[][Section 7]{valori16}, and so our results should not be affected by such a change.

The modelling of the magnetic field enabled us to follow the evolution of the relative magnetic helicity and its two components, the non-potential, and the volume-threading one, in AR 12673. The negative sign of all helicity components confirmed the findings of other authors. The values of relative helicity that we estimate are similar in magnitude with those reported by \citet{vemareddy19}, reaching up to $-5\times 10^{43}\,\mathrm{Mx}^2$. They were however three orders of magnitude higher than those of \citet{yan18}, although both cited works used the flux integration method for computing helicity. Compared to \citet{liu18}, our helicity values are again much larger, but these authors approximated helicity only with its twist, and so the difference is reasonable. The change of helicity during the first flare that we found was also much smaller than theirs. 

Additionally, we examined the evolution of the various energy budgets in AR 12673. The evolution pattern of the free energy that we derived seems to agree with \citet{mitra18}, although a direct comparison is not possible since the authors provide only the ratio $E_\mathrm{f}/E_\mathrm{p}$. The field energy that we derive, which is as high as $2.5\times 10^{33}\,\mathrm{erg}$, compares also well with that of \citet{vemareddy19} that is computed as the sum of the accumulated energy in the system.

The focus of this work was the examination of the behaviour of the helicity ratio $\lvert H_\mathrm{j}\rvert/\lvert H\rvert$ during the two flares. This first observational determination of the evolution of this ratio in an AR seems to confirm the findings of \citet{pariat17}; the helicity ratio increases before major flares, and it relaxes afterwards. Moreover, the much smaller current-carrying part of helicity, i.e., the fact that $\lvert H_\mathrm{j}\rvert/\lvert H\rvert < 1$, is in agreement with previous results \citep{moraitis14}. Also, the examination of the energy ratio $E_\mathrm{f}/E$ is compatible with the results of \citet{pariat17} and \citet{zuccarello18}; this ratio relates also to AR eruptivity, but to a smaller extent than $\lvert H_\mathrm{j}\rvert/\lvert H\rvert$, and without the capability of providing a threshold above which flares occur, unlike the helicity ratio. 

In order to establish these results, the evolution of the helicity ratio should be examined in a large number of ARs, with different characteristics regarding their evolutionary stage and/or their eruptivity, and with the highest possible cadence. A step towards this direction is made with the work of Thalmann et al. (in preparation). The present paper nevertheless, provides the first observational support to the relation of the helicity ratio $\lvert H_\mathrm{j}\rvert/\lvert H\rvert$ with solar eruptivity.

\begin{acknowledgements}
E. Pariat, L. Linan, and K. Moraitis acknowledge the support of the French Agence Nationale pour la Recherche through the HELISOL project, contract n$^\mathrm{o}$ ANR-15-CE31-0001. X. Sun is partially supported by NSF award n$^\mathrm{o}$ 1848250. This work was supported by the Programme National PNST of CNRS/INSU co-funded by CNES and CEA.
\end{acknowledgements}

\bibliographystyle{aa}
\bibliography{refs}

\begin{thebibliography}{39}
\expandafter\ifx\csname natexlab\endcsname\relax\def\natexlab#1{#1}\fi

\bibitem[{{Augusto} {et~al.}(2019){Augusto}, {Navia}, {de Oliveira},
  {Nepomuceno}, {Fauth}, {Kopenkin}, \& {Sinzi}}]{augusto19}
{Augusto}, C.~R.~A., {Navia}, C.~E., {de Oliveira}, M.~N., {et~al.} 2019,
  Publications of the Astronomical Society of the Pacific, 131, 024401

\bibitem[{{Berger}(1999)}]{berger99}
{Berger}, M.~A. 1999, Plasma Physics and Controlled Fusion, 41, B167

\bibitem[{{Berger} \& {Field}(1984)}]{BergerF84}
{Berger}, M.~A. \& {Field}, G.~B. 1984, J. Fluid. Mech., 147, 133

\bibitem[{{Chertok} {et~al.}(2018){Chertok}, {Belov}, \& {Abunin}}]{chertok18}
{Chertok}, I.~M., {Belov}, A.~V., \& {Abunin}, A.~A. 2018, Space Weather, 16,
  1549

\bibitem[{{Finn} \& {Antonsen}(1985)}]{fa85}
{Finn}, J. \& {Antonsen}, T. 1985, Comments Plasma Phys. Control. Fusion, 9,
  111

\bibitem[{{Green} {et~al.}(2018){Green}, {T{\"o}r{\"o}k}, {Vr{\v{s}}nak},
  {Manchester}, \& {Veronig}}]{green18}
{Green}, L.~M., {T{\"o}r{\"o}k}, T., {Vr{\v{s}}nak}, B., {Manchester}, W., \&
  {Veronig}, A. 2018, \ssr, 214, 46

\bibitem[{{Hou} {et~al.}(2018){Hou}, {Zhang}, {Li}, {Yang}, \& {Li}}]{hou18}
{Hou}, Y.~J., {Zhang}, J., {Li}, T., {Yang}, S.~H., \& {Li}, X.~H. 2018, \aap,
  619, A100

\bibitem[{{Inoue} {et~al.}(2018){Inoue}, {Shiota}, {Bamba}, \&
  {Park}}]{inoue18}
{Inoue}, S., {Shiota}, D., {Bamba}, Y., \& {Park}, S.-H. 2018, \apj, 867, 83

\bibitem[{{James} {et~al.}(2018){James}, {Valori}, {Green}, {Liu}, {Cheung},
  {Guo}, \& {van Driel-Gesztelyi}}]{james18}
{James}, A.~W., {Valori}, G., {Green}, L.~M., {et~al.} 2018, \apjl, 855, L16

\bibitem[{{Linan} {et~al.}(2018){Linan}, {Pariat}, {Moraitis}, {Valori}, \&
  {Leake}}]{linan18}
{Linan}, L., {Pariat}, {\'E}., {Moraitis}, K., {Valori}, G., \& {Leake}, J.
  2018, \apj, 865, 52

\bibitem[{{Liu} {et~al.}(2018){Liu}, {Cheng}, {Wang}, {Zhou}, {Guo}, \&
  {Cui}}]{liu18}
{Liu}, L., {Cheng}, X., {Wang}, Y., {et~al.} 2018, \apj, 867, L5

\bibitem[{{Mitra} {et~al.}(2018){Mitra}, {Joshi}, {Prasad}, {Veronig}, \&
  {Bhattacharyya}}]{mitra18}
{Mitra}, P.~K., {Joshi}, B., {Prasad}, A., {Veronig}, A.~M., \&
  {Bhattacharyya}, R. 2018, \apj, 869, 69

\bibitem[{{Moraitis} {et~al.}(2014){Moraitis}, {Tziotziou}, {Georgoulis}, \&
  {Archontis}}]{moraitis14}
{Moraitis}, K., {Tziotziou}, K., {Georgoulis}, M.~K., \& {Archontis}, V. 2014,
  \solphys, 289, 4453

\bibitem[{{Morosan} {et~al.}(2019){Morosan}, {Carley}, {Hayes}, {Murray},
  {Zucca}, {Fallows}, {McCauley}, {Kilpua}, {Mann}, {Vocks}, \&
  {Gallagher}}]{morosan19}
{Morosan}, D.~E., {Carley}, E.~P., {Hayes}, L.~A., {et~al.} 2019, Nature
  Astronomy

\bibitem[{{Nindos} \& {Andrews}(2004)}]{nindos04}
{Nindos}, A. \& {Andrews}, M.~D. 2004, \apjl, 616, L175

\bibitem[{{Pariat} {et~al.}(2017){Pariat}, {Leake}, {Valori}, {Linton},
  {Zuccarello}, \& {Dalmasse}}]{pariat17}
{Pariat}, {\'E}., {Leake}, J.~E., {Valori}, G., {et~al.} 2017, \aap, 601, A125

\bibitem[{{Pariat} {et~al.}(2015){Pariat}, {Valori}, {D{\'e}moulin}, \&
  {Dalmasse}}]{pariat15}
{Pariat}, {\'E}., {Valori}, G., {D{\'e}moulin}, P., \& {Dalmasse}, K. 2015,
  \aap, 580, A128

\bibitem[{{Pesnell} {et~al.}(2012){Pesnell}, {Thompson}, \&
  {Chamberlin}}]{pes12}
{Pesnell}, W.~D., {Thompson}, B.~J., \& {Chamberlin}, P.~C. 2012, \solphys,
  275, 3

\bibitem[{{Rust}(1994)}]{rust94}
{Rust}, D.~M. 1994, \grl, 21, 241

\bibitem[{{Scherrer} {et~al.}(2012){Scherrer}, {Schou}, {Bush}, {Kosovichev},
  {Bogart}, {Hoeksema}, {Liu}, {Duvall}, {Zhao}, {Title}, {Schrijver},
  {Tarbell}, \& {Tomczyk}}]{sche12}
{Scherrer}, P.~H., {Schou}, J., {Bush}, R.~I., {et~al.} 2012, \solphys, 275,
  207

\bibitem[{{Sun} \& {Norton}(2017)}]{sun17}
{Sun}, X. \& {Norton}, A.~A. 2017, Research Notes of the American Astronomical
  Society, 1, 24

\bibitem[{{Taylor}(1974)}]{taylor74}
{Taylor}, J.~B. 1974, Physical Review Letters, 33, 1139

\bibitem[{{Thalmann} {et~al.}(2019){Thalmann}, {Linan}, {Pariat}, \&
  {Valori}}]{thalmann19}
{Thalmann}, J.~K., {Linan}, L., {Pariat}, E., \& {Valori}, G. 2019, arXiv
  e-prints, arXiv:1907.01179

\bibitem[{{Tzio\-tziou} {et~al.}(2012){Tzio\-tziou}, {Georgoulis}, \&
  {Raouafi}}]{tgr12}
{Tzio\-tziou}, K., {Georgoulis}, M.~K., \& {Raouafi}, N.-E. 2012, \apjl, 759,
  L4

\bibitem[{{Tziotziou} {et~al.}(2014){Tziotziou}, {Moraitis}, {Georgoulis}, \&
  {Archontis}}]{tziotziou14}
{Tziotziou}, K., {Moraitis}, K., {Georgoulis}, M.~K., \& {Archontis}, V. 2014,
  \aap, 570, L1

\bibitem[{{Valori} {et~al.}(2012){Valori}, {D{\'e}moulin}, \& {Pariat}}]{val12}
{Valori}, G., {D{\'e}moulin}, P., \& {Pariat}, {\'E}. 2012, \solphys, 278, 347

\bibitem[{{Valori} {et~al.}(2013){Valori}, {D{\'e}moulin}, {Pariat}, \&
  {Masson}}]{val13}
{Valori}, G., {D{\'e}moulin}, P., {Pariat}, {\'E}., \& {Masson}, S. 2013, \aap,
  553, A38

\bibitem[{{Valori} {et~al.}(2016){Valori}, {Pariat}, {Anfinogentov}, {Chen},
  {Georgoulis}, {Guo}, {Liu}, {Moraitis}, {Thalmann}, \& {Yang}}]{valori16}
{Valori}, G., {Pariat}, {\'E}., {Anfinogentov}, S., {et~al.} 2016, \ssr, 201,
  147

\bibitem[{{Vemareddy}(2019)}]{vemareddy19}
{Vemareddy}, P. 2019, \apj, 872, 182

\bibitem[{{Verma}(2018)}]{verma18}
{Verma}, M. 2018, \aap, 612, A101

\bibitem[{{Veronig} {et~al.}(2018){Veronig}, {Podladchikova}, {Dissauer},
  {Temmer}, {Seaton}, {Long}, {Guo}, {Vr{\v s}nak}, {Harra}, \&
  {Kliem}}]{veronig18}
{Veronig}, A.~M., {Podladchikova}, T., {Dissauer}, K., {et~al.} 2018, \apj,
  868, 107

\bibitem[{{Wang} {et~al.}(2018){Wang}, {Yurchyshyn}, {Liu}, {Ahn}, {Toriumi},
  \& {Cao}}]{wang18}
{Wang}, H., {Yurchyshyn}, V., {Liu}, C., {et~al.} 2018, Research Notes of the
  American Astronomical Society, 2, 8

\bibitem[{{Wheatland} {et~al.}(2000){Wheatland}, {Sturrock}, \&
  {Roumeliotis}}]{wheatland00}
{Wheatland}, M.~S., {Sturrock}, P.~A., \& {Roumeliotis}, G. 2000, \apj, 540,
  1150

\bibitem[{{Wiegelmann}(2004)}]{wieg04}
{Wiegelmann}, T. 2004, \solphys, 219, 87

\bibitem[{{Wiegelmann} {et~al.}(2006){Wiegelmann}, {Inhester}, \&
  {Sakurai}}]{wieg06}
{Wiegelmann}, T., {Inhester}, B., \& {Sakurai}, T. 2006, \solphys, 233, 215

\bibitem[{{Wiegelmann} {et~al.}(2012){Wiegelmann}, {Thalmann}, {Inhester},
  {Tadesse}, {Sun}, \& {Hoeksema}}]{wieg12}
{Wiegelmann}, T., {Thalmann}, J.~K., {Inhester}, B., {et~al.} 2012, \solphys,
  281, 37

\bibitem[{{Yan} {et~al.}(2018){Yan}, {Wang}, {Pan}, {Kong}, {Xue}, {Yang},
  {Li}, \& {Feng}}]{yan18}
{Yan}, X.~L., {Wang}, J.~C., {Pan}, G.~M., {et~al.} 2018, \apj, 856, 79

\bibitem[{{Yang} {et~al.}(2017){Yang}, {Zhang}, {Zhu}, \& {Song}}]{yang17}
{Yang}, S., {Zhang}, J., {Zhu}, X., \& {Song}, Q. 2017, \apj, 849, L21

\bibitem[{{Zuccarello} {et~al.}(2018){Zuccarello}, {Pariat}, {Valori}, \&
  {Linan}}]{zuccarello18}
{Zuccarello}, F.~P., {Pariat}, E., {Valori}, G., \& {Linan}, L. 2018, \apj,
  863, 41

\end{thebibliography}

\end{document}